# Towards Universal Non-Volatile Resistance Switching in Non-metallic Monolayer Atomic Sheets

Ruijing Ge*[1], Xiaohan Wu*[1], Myungsoo Kim[1], Harry Chou[1], Sushant Sonde[2,3], Li Tao[1,4], Jack C. Lee[1] and Deji Akinwande[1]

[1]Microelectronics Research Center, The University of Texas at Austin, Texas 78758, USA.

[2]Institute for Molecular Engineering, University of Chicago, Chicago, IL, USA.

[3]Center for Nanoscale Materials, Argonne National Laboratory, Lemont, IL, USA.

[4]School of Materials Science and Engineering, Southeast University, Nanjing, 211189, China.

*Equal Contribution.

**Over the past decade, two-dimensional (2D) atomic sheets have inspired new ideas in nanoscience and nanotechnology[1] including topologically-protected charge transport[2,3], spatially-separated excitons[4], and strongly anisotropic heat transport[5]. Here, we report the intriguing observation of stable non-volatile resistance switching (NVRS) in single-layer atomic sheets sandwiched between metal electrodes. NVRS is observed in the prototypical semiconducting ($MX_2$, M=Mo, W; and X=S, Se) transitional metal dichalcogenides (TMDs)[6], and insulating hexagonal boron nitride (h-BN)[7], which alludes to the universality of this phenomenon in non-metallic 2D monolayers, and features forming-free switching. This observation of NVRS phenomenon, widely attributed to ionic diffusion, filament and interfacial redox in bulk oxides and electrolytes[8-11], inspires new studies on defects, ion transport and energetics at the sharp interfaces between atomically-thin sheets and conducting electrodes. From a contemporary perspective, switching is all the more unexpected in monolayers since leakage current is a fundamental limit in ultra-thin oxides[12]. Emerging device concepts in non-volatile flexible memory fabrics, and brain-inspired (neuromorphic) computing could benefit substantially from the pervasive NVRS effect and the associated wide materials and engineering co-design opportunity. Experimentally, a 50 GHz radio-frequency (RF) monolayer switch is demonstrated, which opens up a new application for electronic zero-static power RF switching technology.**

**Main**

Recently, non-volatile resistance switching has been observed in various solution-processed multi-layer 2D material morphologies including pristine, reduced and functionalized graphene oxide mixtures, partially degraded black phosphorus, functionalized MoS$_2$ and composites, and TMD based hybrids[13-16], where the resistance can be modulated between a high-resistance state (HRS) and a low-resistance state (LRS) and subsequently retained absent any power supply. In addition, several researchers have reported NVRS in multi-layer h-BN[17, 18]. However, it was believed that NVRS phenomenon was not accessible in single-layer atomic sheets[13, 19], likely due to excessive leakage current that is known to prevent sub-nanometre scaling in conventional oxide-based vertical metal-insulator-metal (MIM) configuration[12]. Sangwan et al., circumvented this issue with the discovery that atomic rearrangement of certain grain boundaries in single-layer MoS$_2$ produced NVRS in lateral devices[20]. However, for practical nanotechnology, vertical MIM devices are of contemporary preference owing to their smaller footprint and denser integration.

In this article, we report our observation of NVRS phenomenon in a variety of single-layer non-metallic atomic sheets (MoS$_2$, MoSe$_2$, WS$_2$, WSe$_2$ and h-BN) in desired MIM configuration, overcoming a fundamental limitation in ultra-scaled bulk materials. These devices can be collectively labelled 2D non-volatile switches (2DNS). Furthermore, controlled experiments with active (Ag), noble (Au), and graphene electrodes, and on a single-crystalline device also produce NVRS, suggesting a rich multi-physics mechanism at the atomic limit. Moreover, the rapid commercialization of large-area synthesis and the ever-increasing portfolio of 2D nanomaterials and alloys, indicate that atomic sheets can play a substantial role for numerous emerging applications such as non-volatile switches and memory fabrics, memristor circuits and computing.

Synthetic atomic sheets prepared by standard chemical vapor deposition (CVD)[21] and metal-organic CVD[22] and subsequently transferred (**Supplementary Fig. 1**) onto device substrates, were investigated for NVRS effect in MIM configuration as illustrated in **Fig. 1a-c**. Optical images of so-called crossbar and lithography-free MIM devices are shown in **Fig. 1d,e,** respectively. To avoid effects due to metal oxides, inert metal, gold, is used as electrodes (if not specified) and thus ensure that 2D materials play the active role in the resistive switching behaviour. Cross-sectional transmission electron microscopy (TEM) was conducted to evaluate a representative fabricated MIM sandwich (**Fig. 1f**), revealing a sharp and clean interface. Raman and photoluminescence spectroscopy were used to characterize the $MX_2$ sheets showing clear evidence of single-layer spectra as displayed in **Fig. 1g,h**, respectively. Similarly, Raman and x-ray photoelectron spectroscopy were used to evaluate CVD h-BN[23] (**Fig. 1i** and **Supplementary Fig. 2**).

DC electrical measurements were performed on as-fabricated crossbar devices consisting of atomic sheets with Au bottom and top electrodes, and revealed non-volatile resistance switching in the monolayer $MX_2$ and h-BN active layers (**Fig. 2**). For instance, $MoS_2$, the prototypical TMD, featured low currents corresponding to a high-resistance state until the application of ~1V, which SET the atomic-layer switch to a low-resistance state that persists until a negative voltage is applied to RESET it (**Fig. 2a**). Absent electrical power, this atomic switch consumes zero-static power and affords data retention under ambient conditions (**Supplementary Fig. 3**) for more than 100x longer than previously reported lateral counterpart[20]. Interestingly, the single-layer non-volatile switch required no electro-forming step, a prerequisite in transition metal oxides (TMOs) that initializes a soft dielectric breakdown to form a conductive filament for subsequent NVRS operation[8, 9]. Although it has been shown that electroforming can be avoided with thickness scaling into the nm-regime, excessive leakage current from trap-assisted tunnelling is a limiting consequence[9, 12]. Here, an ON/OFF ratio above $10^4$ can be achieved, highlighting a defining advantage of crystalline monolayers over ultrathin amorphous oxides.

Low-voltage 'READ' characteristics of the HRS and LRS exhibit a linear profile over a small voltage range (**Fig. 2b,c**), though the magnitudes of the current can differ by a ratio of more than $10^4$. While the 'READ' profiles appears superficially similar, temperature dependent studies over a larger voltage range indicate uniquely different transport mechanisms in the two states (Schottky emission in HRS and localized direct tunnelling in LRS) and will be discussed further with respect to **Fig. 4**.

Certain single-layer $MoS_2$ devices of the same MIM construction feature unipolar switching where voltage of the same polarity is used for both SET and RESET programming (**Fig. 2d**). Motivated by the observation of NVRS in $MoS_2$, the quartet of single-layer $MX_2$ was investigated and all showed similar intriguing characteristics of (predominantly) bipolar switching, and (occasional) unipolar switching as demonstrated in **Fig. 2e-j**. Similar qualitative results were achieved with h-BN (**Fig. 2k,l**). These collective results of NVRS in representative atomic sheets allude to a universal effect in non-metallic 2D nanomaterials which opens a new avenue of scientific research on defects, charge, and interfacial phenomena at the atomic scale, and the associated materials design for diverse applications. Regarding the polarity dependence, the precise understanding of the factors that produce either bipolar or unipolar switching in monolayer sheets is yet unclear and deserving of atomistic modelling and microscopy studies for elucidation. A recent study in TMOs have suggested that the co-existence of bipolar and unipolar switching is a complex competition among several parameters including lateral area, grain size, and compliance conditions[24]. Nevertheless, the underlying physics of unipolar switching has been previously established to be due to electro-thermal heating that facilitates diffusion[8,9]. A symptom of this effect is the relatively higher RESET current needed to increase the local temperature to break the conductive link[9].

In the majority of experiments, Au was selected as an inert electrode to rule out any switching effect that might arise from interfacial metal oxide formation. Indeed, devices of identical processing with Au electrodes using Cr as an adhesion layer (for the lift-off of TE), but without a TMD active-layer, did not reveal NVRS (**Supplementary Fig. 4**). Furthermore, to rule out the undesirable contribution of polymer contamination from microfabrication, very clean devices including lithography-free and transfer-free devices, and litho-free and transfer resist-free $MoS_2$ devices (**Fig. 3a**) were made, which also produced the NVRS effect, alluding to an intrinsic origin. The lithography-free and transfer-free devices are based on monolayer $MoS_2$ grown on gold foil[25] or monolayer h-BN grown on nickel foil (**Supplementary Fig. 5b**). Then gold top electrode is directly deposited using e-beam evaporation via laser shadow mask. Thus, no transfer process or lithography process is used, excluding possible residues from transfer and lithography. The litho-free and transfer resist-free process (**Supplementary Fig. 1c**) was specially developed to prevent the direct contact between as-grown $MoS_2$ surface and polymers that can leave residues.

For 1L-$MoS_2$ switches with pure Au electrodes, HRS tunnel resistance is typically ~0.1-10 $M\Omega$-$\mu m^2$, which is within estimates from local probe measurements[26]. It has been previously reported that line or grain boundary defects in polycrystalline 2D multi-layers play an intrinsic role in switching[27]. While a possible factor in monolayers, it is not an exclusive factor as shown in **Fig. 3b** from an MIM device realized on a single-crystal (boundary-free) CVD MoS$_2$ (**Supplementary Fig. 5c**), highlighting the potential role of localized effects. Also, the phenomenon is not restricted to inert electrodes since monolayer TMD with electrochemically active (Ag) electrodes[8, 14], produce NVRS as presented in **Fig. 3c** (see **Supplementary Fig. 5** for devices with Ni and Cr electrodes). Moreover, monolayer graphene is also a suitable electrode option (**Fig. 3d**). Interestingly, these results open up the design space for electrode engineering (work-function and interface redox) from inert to active metals to 2D semimetals, the latter offering the potential of atomically-thin sub-nanometre 2DNS for ultra-flexible and dense non-volatile computing fabrics.

In order to gain insight into the underlying mechanism(s), electrical measurements over five degrees of freedom, namely, temperature, area scaling, compliance current, voltage sweep rate and layer thickness were conducted using MoS$_2$ as the active layer owing to its greater material maturity. The I-V characteristics at different temperatures are analysed to explain the electron transport mechanisms at LRS and HRS. At LRS (**Fig. 4a**), metallic Ohmic conduction can be deduced since the current decreases as the temperature increases, and the normalized conductance $G_n = (dI/dV)/(I/V)$ is approximately unity, a signature of linear transport attributed to direct tunnelling, $J \propto KV \exp\left(\frac{-4\pi d\sqrt{2m^*\varphi}}{h}\right)$, where $J$ is the current density, $m^*$ is the effective mass, $\varphi$ is the tunnel barrier height, $h$ is Planck's constant, and $K$ is proportional to the lateral area ($A$) and dependent on the barrier parameters ($m$, $\varphi$, $d$)[28]. $d$ is the 2D barrier thickness. The direct tunnelling model produces linear transport and is consistent with an MIM band diagram (**Fig. 4a**). At HRS (**Fig. 4b**), non-linear I-V characteristics are observed, with the current increasing as the temperature increases. The HRS data was best-fitted by the Schottky emission model[28] with good agreement (**Fig. 4c**); $J \propto A^* T^2 \exp\left[\frac{-q(\phi_B - \sqrt{qE/4\pi\varepsilon_r\varepsilon_0})}{kT}\right]$, $A^* = \frac{120 m^*}{m_0}$, where $A^*$ is the effective Richardson constant, $m_0$ is the free electron mass, $T$ is the absolute temperature, $q$ is the electronic charge, $\phi_B$ is the Schottky barrier height, $E$ is the electric field across the dielectric, $k$ is Boltzmann's constant, $\varepsilon_0$ is the permittivity in vacuum, and $\varepsilon_r$ is the optical dielectric constant. The effective thickness of ~1nm is used and $m^*/m_0$ is ~1. The extracted barrier height is ~0.47eV at 300K, and the refractive index $\mathbf{n} = \sqrt{\varepsilon_r}$ is 6.84 [29]. Based on the conduction mechanisms, the resistance switching of $MoS_2$ devices can be explained by the proposed model that, in the SET process, the electrons are transported through the conductive pass/filament formed by electrochemical metallization, and in the RESET process, the conductive path is broken, resulting in a Schottky junction at the device interfaces. For h-BN devices, similar metallic Ohmic conduction is seen at LRS characteristics, while Poole-Frenkel emission is the best-fit model for HRS (**Supplementary Fig. 6**), indicating a different mechanism.

Area scaling studies clearly show distinct profiles with the LRS relatively flat while the HRS features a more complex relation (**Fig. 4d**). The LRS profile is consistent with the theory of a single (or few) localized filament(s)[8, 9, 11]. Below 100 $\mu m^2$, the HRS resistance scales inversely with area due to uniform conduction. For larger sizes, the resistance becomes area-invariant and is attributed to the presence of localized grain boundaries. We note that the average domain size of typical CVD $MoS_2$ monolayer is ~$10^2$-$10^3$ $\mu m^2$. The SET/RESET voltage - area scaling is presented in **Supplementary Fig. 7**. The current and resistance dependence on compliance current (**Fig. 4e,f**) reveal a linear scaling that can be credited to an increase in the cross-sectional area of a single filament or to the formation of multiple filaments[11]. For applications, such programmable resistance states are suitable for multilevel memory. In addition, the intrinsic low-resistance values, approaching ~5 $\Omega$ (**Fig. 4f**), open up new opportunities for nanoscale low-power non-volatile RF switches. The dependence of the SET/RESET voltages on sweep rate (**Fig. 4g**) suggests that slower rates afford more time for ionic diffusion resulting in reduced voltages, an important consideration for low-voltage operation. Layer dependent studies up to four layers (**Supplementary Fig. 8**) demonstrate that the switching phenomena persists (**Fig. 4h**), with a distinction that the LRS resistance increases with layer number (**Fig. 4i**).

Given the worldwide interest in layered atomic materials[1, 5], it is important to consider potential applications. One emerging target is next-generation solid-state memory, with phase-change and TMO devices currently the leading candidates over the past decade. 2DNS devices offer distinct advantages in terms of ultimate vertical scaling, down to an atomic layer with forming-free operation. With the replacement of metal electrodes with graphene, the entire memory cell can be scaled below 2 nm. Moreover, the transparency of graphene and the unique spectroscopic features of 2D materials affords direct optical characterization for in-situ studies and in-line manufacturing testing. Presently, manual endurance data (**Fig. 5a,b**) is not yet sufficient to meet the stringent requirements for solid-state memory, a reflection of the nascent state of 2DNS compared to TMO memory, which had similar endurance (<$10^3$ cycles) in early research but has now advanced above $10^6$ cycles[9]. Oxygenation by interface engineering or doping may improve endurance, similar to what was observed for amorphous-carbon memory devices[30]. Retention of non-volatile states tested up to a week (**Fig. 5c**) is already sufficient for certain neuromorphic applications involving short and medium term plasticity[31]. Moreover, the sub-nanometre thinness of monolayers is promising for realizing ultra-scaled areas and volume down to 1 $nm^2$ and 1 $nm^3$, respectively. At this limit, a loose 2DNS density of $10^{15}/mm^3$ (10 nm pitch) would provide ample room to mimic the density of human synapses (~$10^9/mm^3$)[32]. For single-bit single-level memory storage, this corresponds to a theoretical areal density of 6.4 Tbit/$in^2$.

Non-volatile low-power RF switches represents another major application of 2DNS. Contemporary switches are realized with transistor or micro-electromechanical devices, both of which are volatile with the latter also requiring large switching voltages, unideal for mobile technologies. Recently, phase-change switches have attracted interest[33], however, the high-temperature melting requirements and slow switching times have limited their utility. In this light, 2DNS offers an unprecedented advancement for high-frequency systems owing to its low voltage operation, small form-factor, fast switching speed, and low-temperature integration compatible with Si or flexible substrates. Initial experimental non-volatile RF switches show promising results with acceptable insertion loss of ~1 dB and isolation of ~12 dB up to 50 GHz (**Fig. 5d**). The extracted ON resistance, $R_{ON}$ ~ 11 Ω and OFF capacitance, $C_{OFF}$ ~ 7.7 fF. This results in a figure of merit (FOM) cut-off frequency, $f_{co}=1/(2\pi R_{ON} C_{OFF})$ ~1.8 THz. The FOM is used for evaluating RF switches[33, 34]. Further advancements especially in terms of scaling is expected to yield significantly higher FOM (**Supplementary Fig. 12**). Importantly, the unique combination of (approximately) area-independent LRS resistance and area-dependent HRS capacitance, yields a FOM that can be scaled to 100s of THz by reduction of device area in sharp contrast to conventional phase-change switches[33, 3433, 3433, 34], where capacitance is proportional to width but $R_{ON}$, inversely dependent, hence, preventing frequency scalability without significantly compromising loss. Furthermore, the high breaking strain and ease of integration of 2D materials on soft substrates[5] can afford flexible non-volatile digital and analog/RF switches that can endure mechanical cycling (**Fig. 5e** and

**Supplementary Fig. 9**).

In summary, we report non-volatile resistance switching in monolayer TMDs (MoS$_2$, MoSe$_2$, WS$_2$, WSe$_2$) and h-BN, alluding to a universal effect in non-metallic 2D nanomaterials. This unexpected result in atomically-thin sheets is likely due to the inherent layered crystalline nature that produces sharp interfaces and clean tunnel barriers, which prevents excessive leakage and affords stable phenomenon so that NVRS can be used for diverse applications in memory fabrics, RF switching, and neuromorphic computing.

**Acknowledgements**

This work was supported in part by Northrop Grumman Aerospace Systems (NGAS). D.A. acknowledges the Presidential Early Career Award for Scientists and Engineers (PECASE). We are grateful to Jesse Tice and Xin Lan of NGAS for collaborative discussion regarding RF switch and design. In addition, we are thankful to Shimeng Yu of Arizona State University, Sabina Spiga of CNR-IMM, Italy, and the group of Philip Wong of Stanford University for useful discussions. We acknowledge the supply of MOCVD samples. We appreciate Maruthi Nagavalliyogeesh and Anupam Roy of the University of Texas at Austin for TEM image rendering and x-ray photoelectron spectroscopy, respectively. We thank the group of Nanshu Lu of the University of Texas at Austin for providing mechanical bending apparatus, and Jo Wozniak of Texas Advanced Computing Centre (TACC) for 3D renderings.

**Author Contributions**

R.G. performed CVD growth, materials transfer, characterization, and device fabrication. H.C. conducted h-BN growth, characterization and transfer. L.T. contributed to sample preparation and device fabrication. S.S. conducted high resolution TEM measurements. X.W. carried out DC electrical measurements of 2D non-volatile resistance switching devices. M.K. conducted RF switch fabrication and measurements. R.G., X.W., J.L. and D.A. analysed the electrical data and characteristics. R.G., X.W. and D.A. initiated the research on cross-plane non-volatile resistance switching in single-layer atomic sheets. All authors contributed to the article based on the draft written by R.G., X.W. and D.A. J.L. and D.A. coordinated and supervised the research.

**Competing Financial Interests**

The authors declare no competing financial interests.

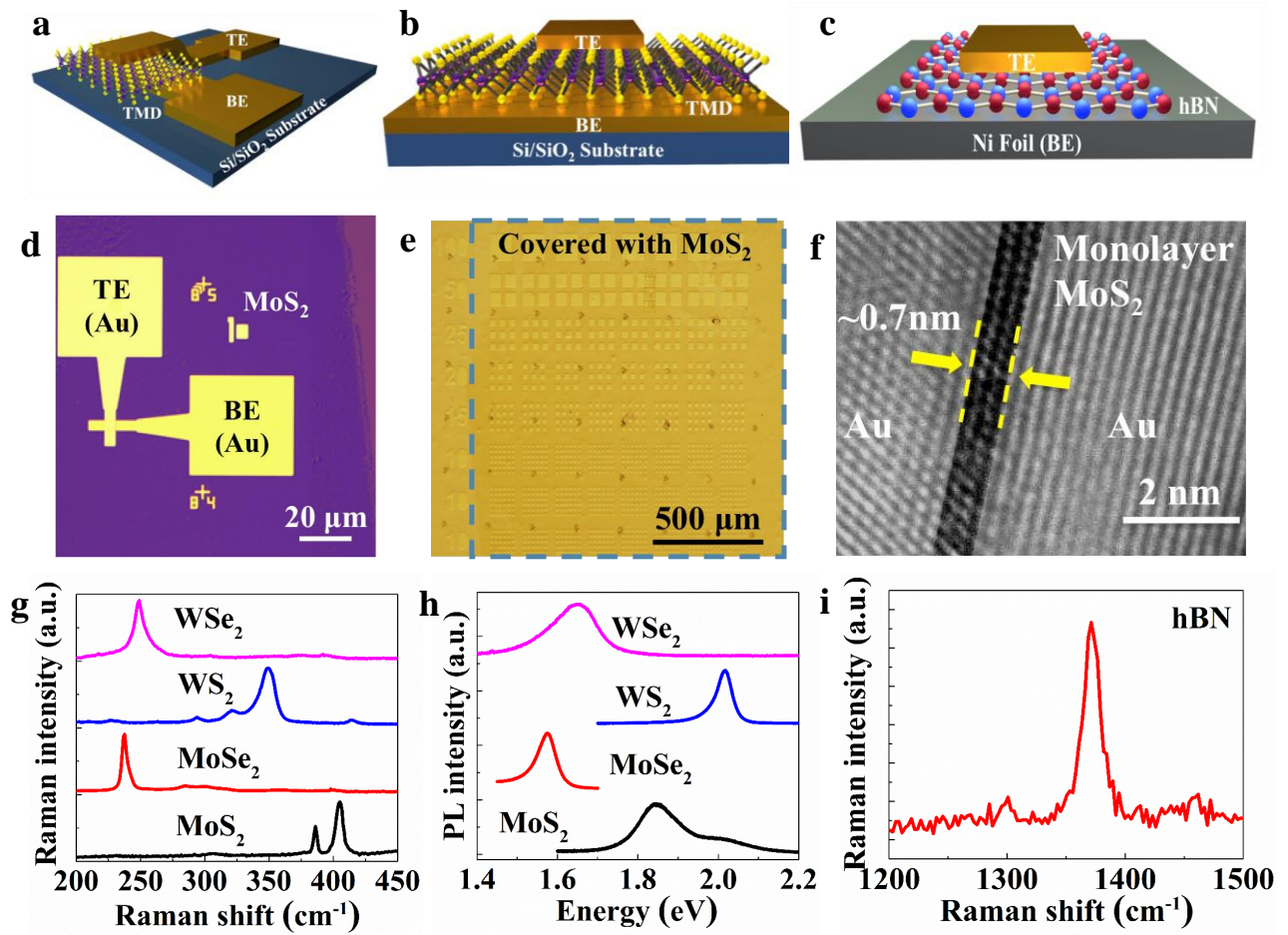

**Figure 1. Schematic, device and material characterization.** The schematics of 2DNS metal-insulator-metal structures including (a) TMD crossbar sandwich, (b) TMD lithography-free sandwich (TE and BE are gold, if not specified), and (c) h-BN transfer-free sandwich (TE is gold). Transferred h-BN MIM devices were also investigated. (d,e) Optical images of fabricated $MoS_2$ crossbar with Au electrodes and litho-free devices on Si/SiO$_2$ substrate. The dash box in (e) indicates the area covered with $MoS_2$. (f) TEM cross-section image of Au/$MoS_2$/Au litho-free device revealing the atomically sharp and clean monolayer interface. (g-h) Raman and PL spectrum of MOCVD-grown monolayer $MoS_2$, $MoSe_2$, $WS_2$ and $WSe_2$, used in the MIM device studies. (i) Raman spectrum of CVD-grown h-BN.

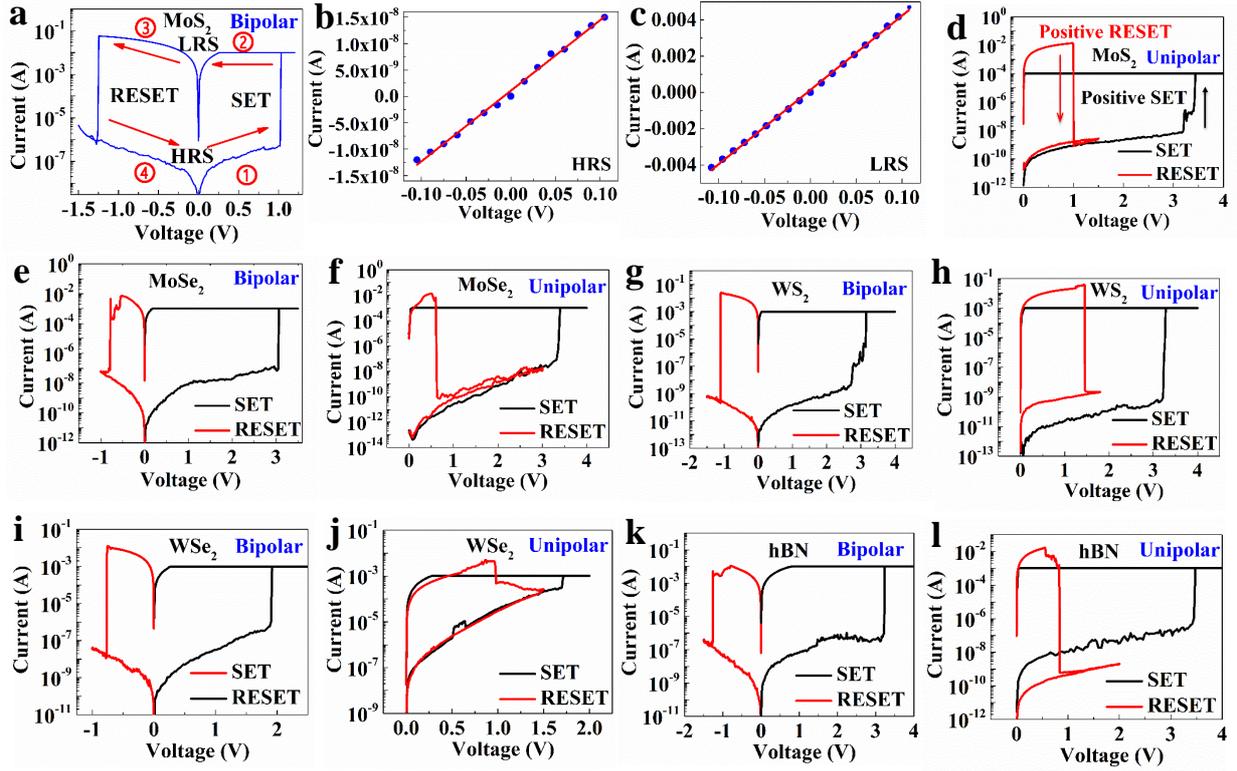

**Figure 2. Typical I-V curves of monolayer TMD and h-BN 2DNS devices.** (a) Representative I-V curve of bipolar resistive switching behaviour in monolayer $MoS_2$ crossbar device with lateral area of $2 \times 2$ μm$^2$. Step 1: voltage increases from 0 to 1.2 V. At ~1V, the current abruptly increases to compliance current, indicating a transition (SET) from high resistance state (HRS) to low resistance state (LRS). Step 2: voltage decreases from 1.2 to 0 V. The device remains in LRS. Step 3: voltage increases from 0 to -1.5 V. At -1.25 V, the current abruptly decreases, indicating a transition (RESET) from LRS to HRS. Step 4: voltage decreases from -1.5 to 0 V. The device remains in HRS until next cycle. (b,c) Low-voltage 'Read' operations at HRS and LRS, confirming the non-volatility. (d) Representative I-V curve of unipolar resistive switching behaviour in monolayer $MoS_2$ crossbar device with lateral area of $2 \times 2$ μm$^2$. For unipolar operation, both SET and RESET transitions are achieved under positive bias. (e-l) Representative I-V curves of bipolar and unipolar resistive switching behaviour in monolayer (e,f) $MoSe_2$, (g,h) $WS_2$, (i,j) $WSe_2$ and (k,l) h-BN crossbar MIM devices, which suggest a universal non-volatile phenomenon in related atomic sheets. The areas of the crossbar devices are $0.4 \times 0.4$ μm$^2$ for $MoSe_2$, $2 \times 2$ μm$^2$ for $WS_2$, $2 \times 2$ μm$^2$ for $WSe_2$, and $1 \times 1$ μm$^2$ for h-BN.

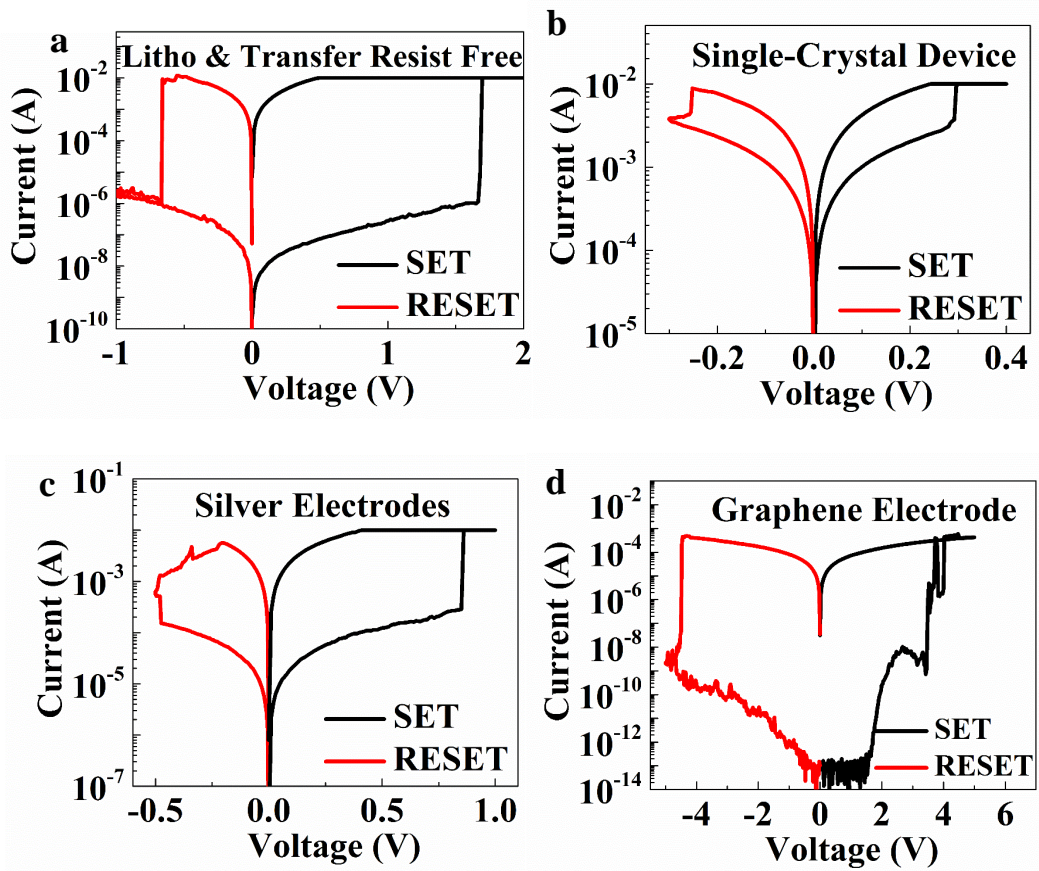

**Figure 3. Typical I-V curves of monolayer MoS$_2$ non-volatile sandwiches with different device conditions.** (a) Representative I-V curve of litho & transfer resist-free Au/MoS$_2$/Au device with lateral area of 15 × 15 μm$^2$. (b) Representative I-V curve of a single-crystal MoS$_2$ device with Au electrodes. The optical image of the device is shown in Supplementary Fig. 5c. (c) Representative I-V curve of a MoS$_2$ litho-free device using silver as top and bottom electrodes. (d) Representative I-V curve of a MoS$_2$ crossbar device using graphene as the top electrode and gold as bottom electrode with an area of 1 × 1 μm$^2$.

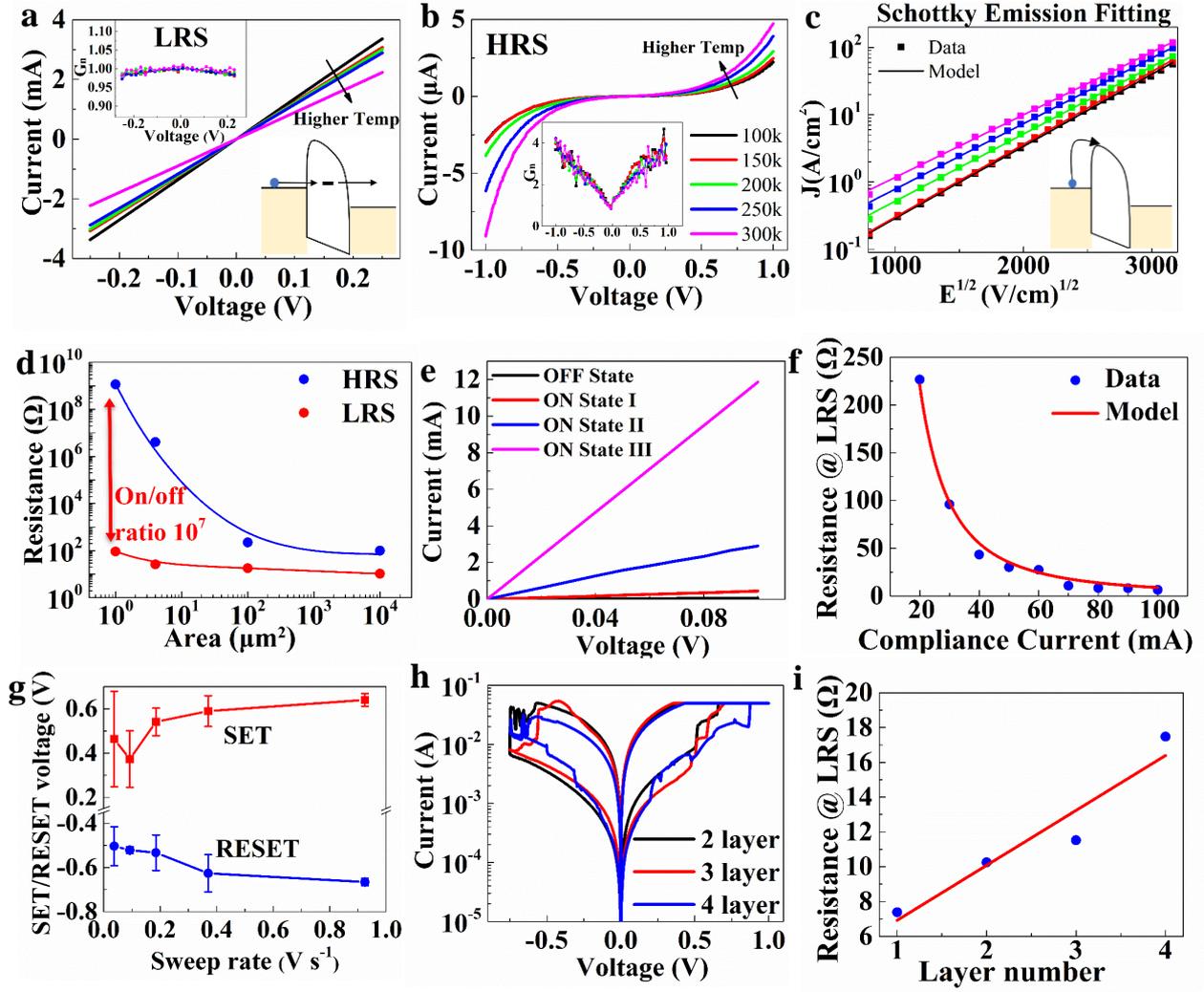

**Figure 4. Dependence of temperature, area scaling, compliance current, sweep rate, and layer thickness of MoS$_2$ non-volatile switching devices.** (a) The I-V characteristics at LRS based on MoS$_2$ crossbar devices at different temperatures indicating a metallic character (temperature legend is in part b). The inset shows the normalized conductance G$_n$ and the schematic of proposed mechanism based on Ohmic direct tunneling via defect(s). (b) The I-V characteristics at HRS at different temperatures. The current increases as the temperature increases. The inset shows the normalized conductance G$_n$. (c) Fitted data using Schottky emission model for HRS. The inset shows the schematic of the model. The device area is 2 x 2 μm$^2$. (d) Area-dependence of low and high resistance states with Au/1L-MoS$_2$/Au structure. The resistances of each state are determined at a low voltage of 0.1 V. The line curves are visual guides. (e) Dependence of the READ current on the compliance current after SET process in MoS$_2$ litho-free 2DNS device. Four separate resistance states (three ON states and one OFF state) are obtained in a single device by varying the compliance current at 20 mA (ON state I), 40mA (ON state II) and 80 mA (ON state III). (f) Relationship between LRS resistance and compliance current indicating a sub-10 Ω resistance is achievable for RF switch applications. The fitting curve is obtained with an inverse quadratic model, $\boldsymbol{y \propto x^{-2}}$. (g) Dependence of the SET and RESET voltages on the sweep rate. The area of this litho-free device is 15 × 15 μm$^2$. (h) Layer dependent I-V characteristics of MoS$_2$ litho-free MIM switches, each with an area of 15 × 15 μm$^2$. (i) Relationship between LRS resistance and layer number of few-layer MoS$_2$ litho-free devices. The straight line is a visual guide. For layer dependent studies, the preparation method for monolayer is MOCVD and PDMS transfer, while few-layer devices are CVD-grown and wet transfer.

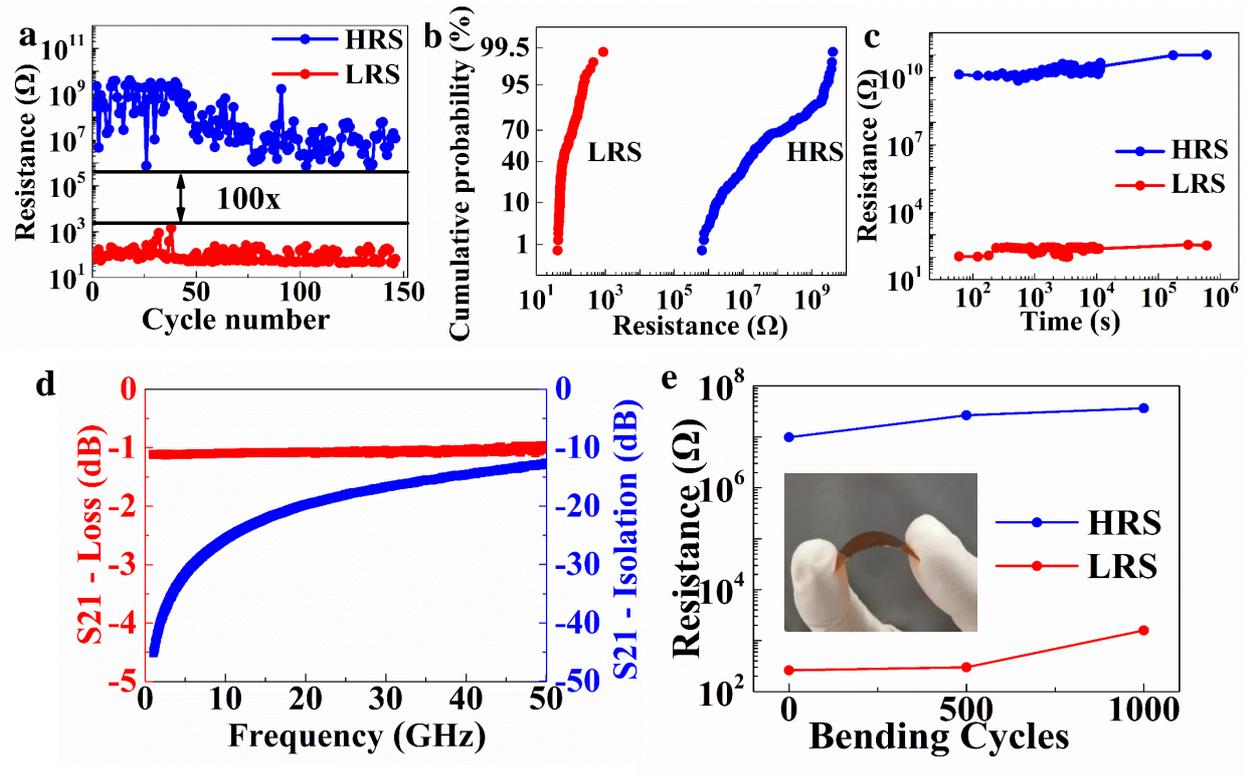

**Figure 5. Representative 2DNS performance.** (a,b) Endurance and resistance distribution of MoS$_2$ crossbar MIM device with 150 manual DC switching cycles. (c) Time dependent measurements of MoS$_2$ crossbar switch featuring stable retention over a week at room temperature. The resistance of the HRS and LRS is determined by measuring the current at a small bias of 0.1 V. The area of this 2L-MoS$_2$ crossbar device is 2 × 2 μm$^2$. (d) Experimental non-volatile RF switches based on 1 × 1 μm$^2$ monolayer MoS$_2$ show promising performance with acceptable insertion loss of ~1 dB and isolation of ~12 dB up to 50GHz. The cutoff frequency figure of merit is ~1.8 THz. (e) Stable resistance of the high-resistance and low-resistance states after 1000 bending cycles at 1% strain.

**Supplementary**

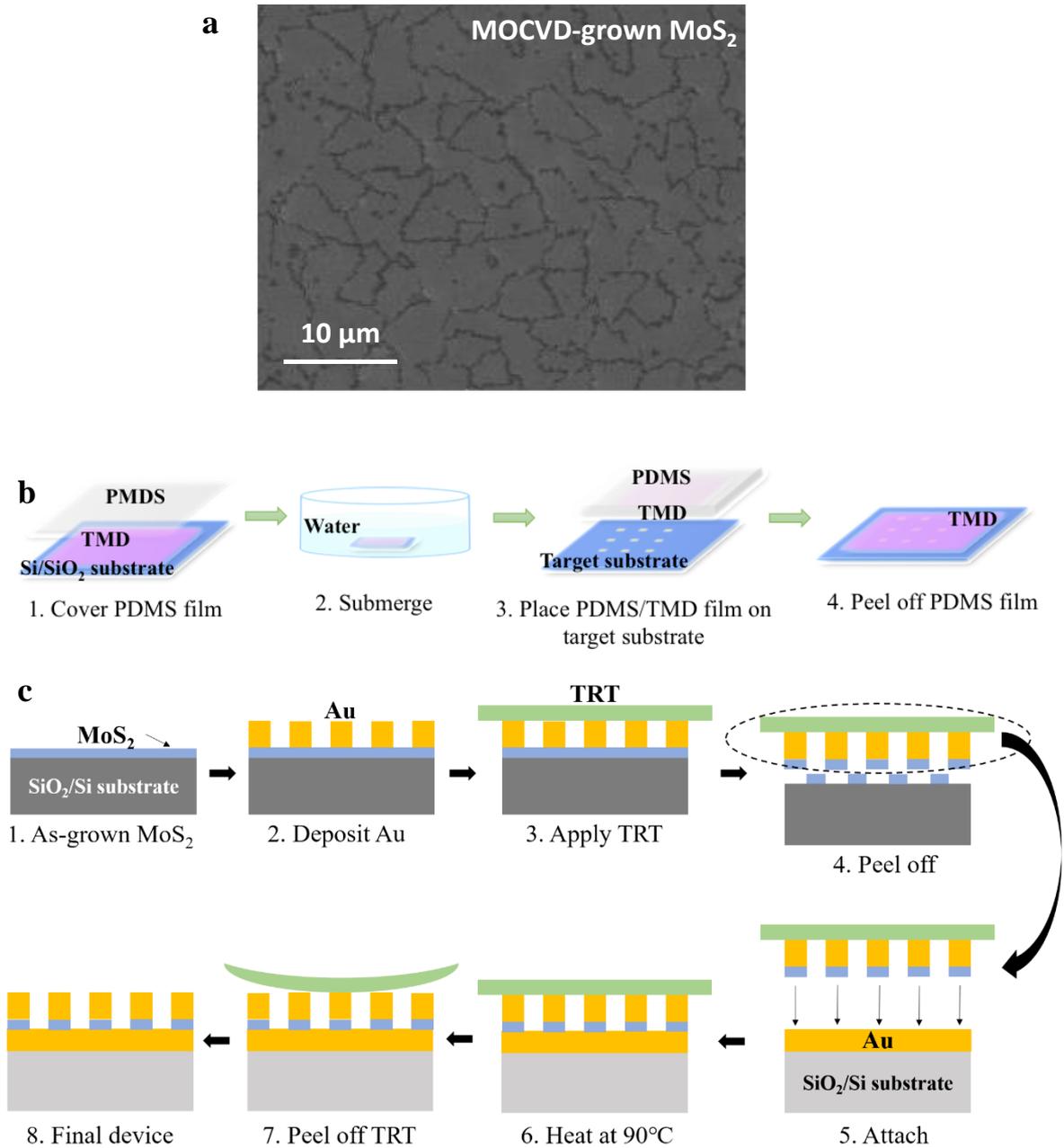

**Supplementary Figure 1.** SEM image and schematics of two TMD transfer processes. (a) SEM image of as-grown MOCVD monolayer MoS$_2$. (b) Simplified illustration of the steps for PDMS assisted pick-and-place transfer fabrication. (c) Illustration of the litho & transfer resist-free process assisted with thermal release tape (TRT). (1) as-grown monolayer MoS$_2$ on SiO$_2$/Si substrate; (2) evaporated Au as top electrode; (3) applying TRT; (4) detaching off TRT/Au/MoS$_2$; (5) integrating TRT/Au/MoS$_2$ onto the target substrate with Au thin film as bottom electrode; (6) heating the target substrate at 90 ℃; (7) peeling off TRT; (8) final completed device on the target substrate. This process avoids contamination that might arise from transfer or lithography.

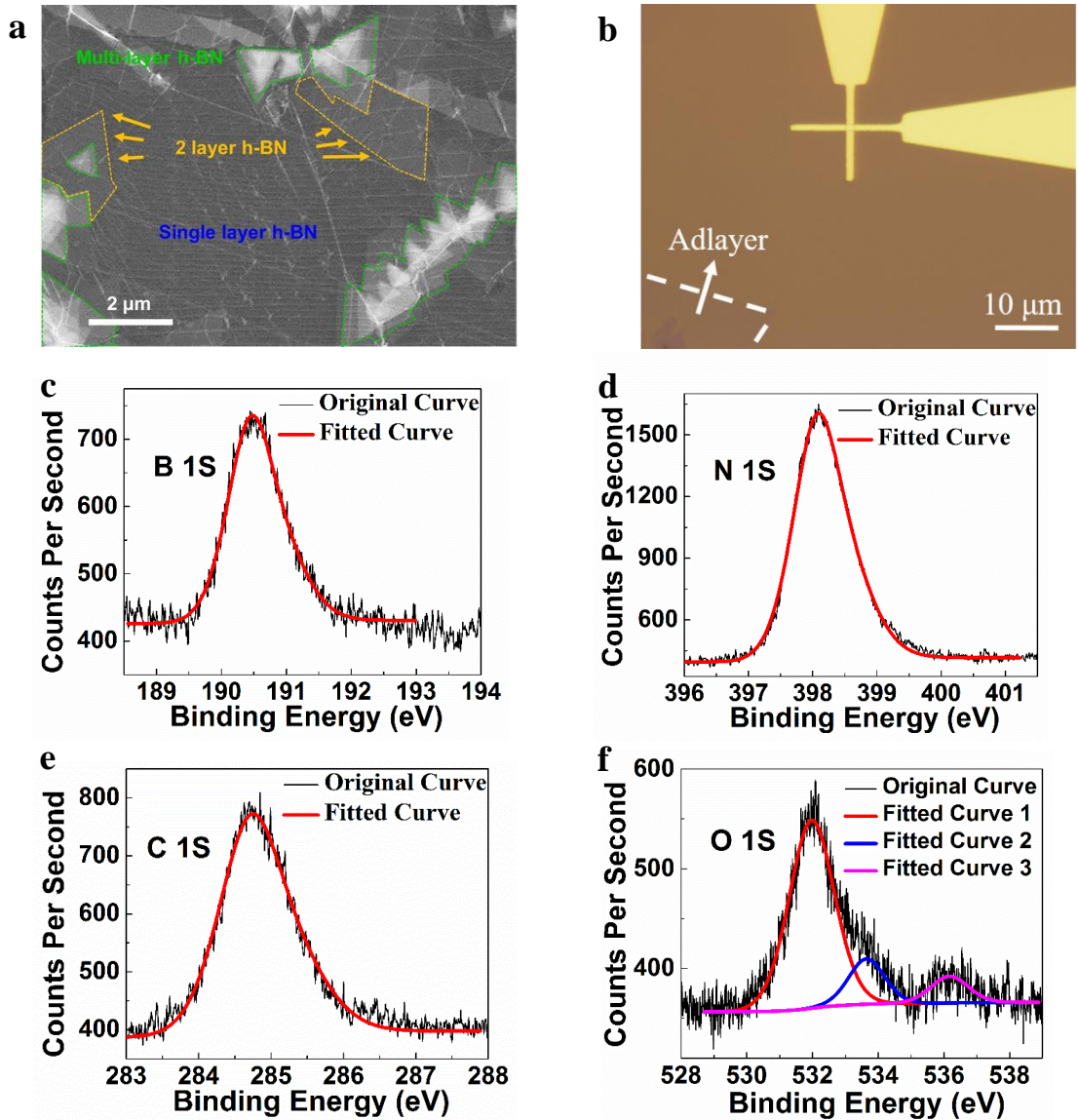

**Supplementary Figure 2. SEM images and XPS of h-BN.** (**a**) SEM image of as-grown CVD h-BN. The coverage of monolayer h-BN is calculated to be 82.8% based on an optical microscope image covering $7\times10^4$ μm$^2$. (**b**) Optical microscope image of an h-BN crossbar device. The dash line section shows the presence of adlayer. The rest of the image is filled with monolayer. (**c-f**) XPS of as-grown h-BN. The B 1s and N 1s peak positions correspond to that which is expected for h-BN at 190.5 eV and 398.1 eV, respectively. The C 1s peak is attributed to C-C bonds from adventitious carbon contamination. The O 1s peak is attributed to C-O bonds also from surface contamination due to ambient conditions.

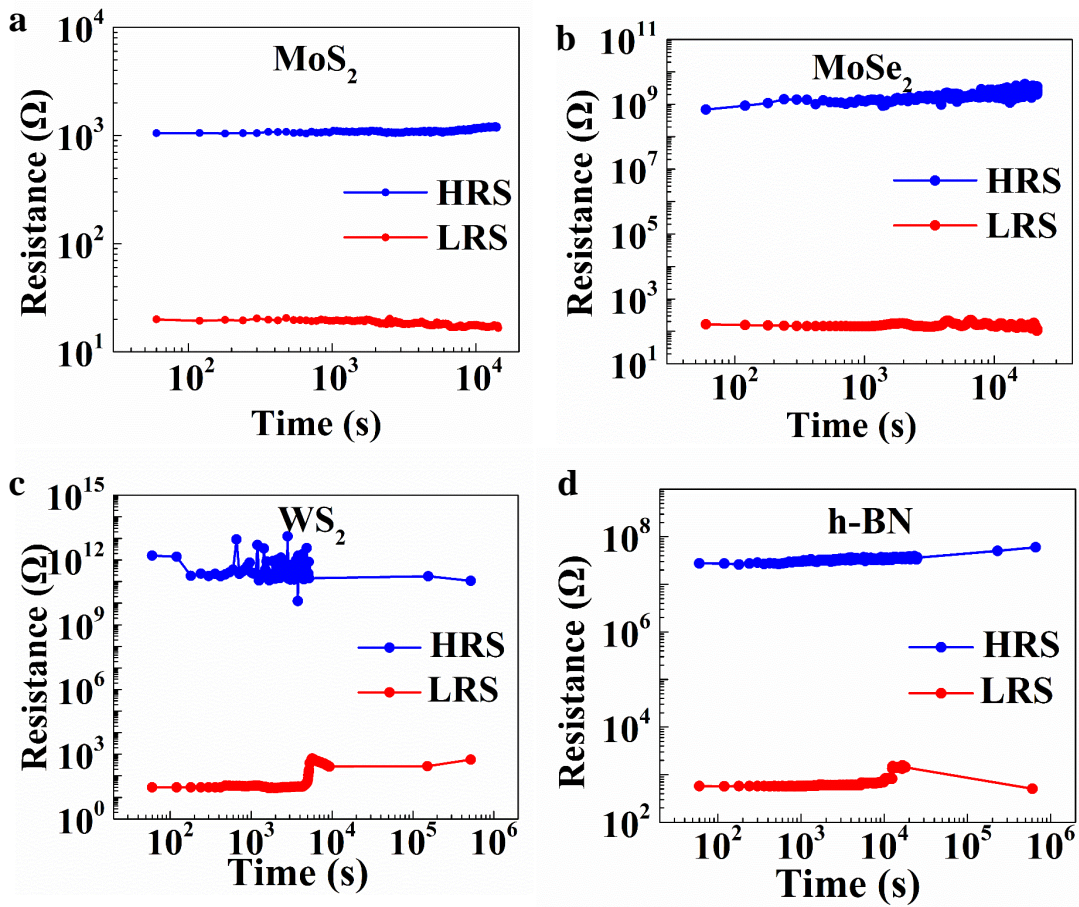

**Supplementary Figure 3.** Retention time of monolayer non-volatile resistance switches. (a-d) The retention time of $MoS_2$, $MoSe_2$, $WS_2$ and h-BN memory devices at room temperature, respectively. The resistance of HRS and LRS is determined by the current read at 0.1 V. The area of the crossbar devices is $10 \times 10$ μm² for $MoS_2$ device, $0.4 \times 0.4$ μm² for $MoSe_2$ device, and $2 \times 2$ μm² for both $WS_2$ and h-BN devices.

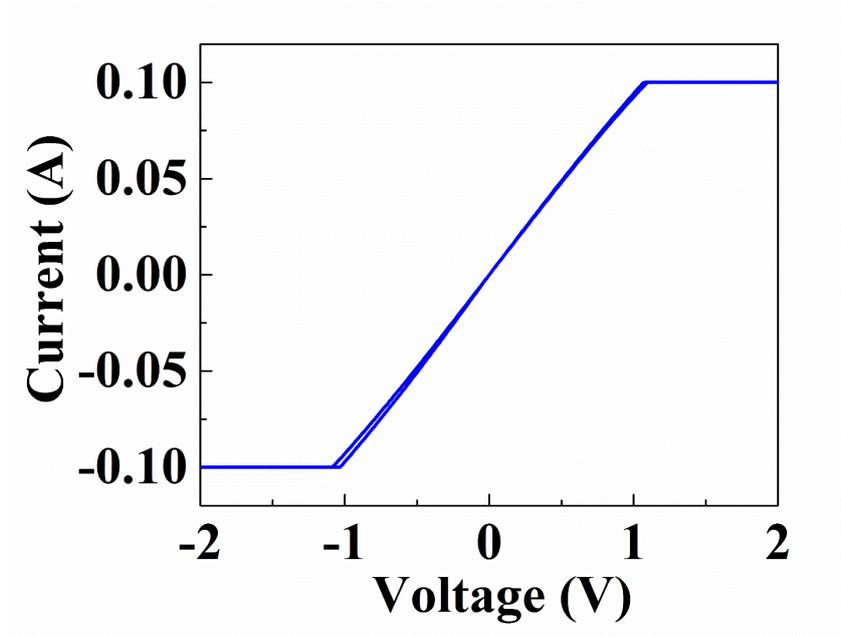

**Supplementary Figure 4.** Test crossbar device without 2D active layer. The structure is Au (60nm) /Cr (2nm) /Au (60nm). Cr serves as the adhesion layer in the lift-off process. No resistive switching behaviour is observed in this structure, indicating that the TMD or h-BN active-layer plays the primary role.

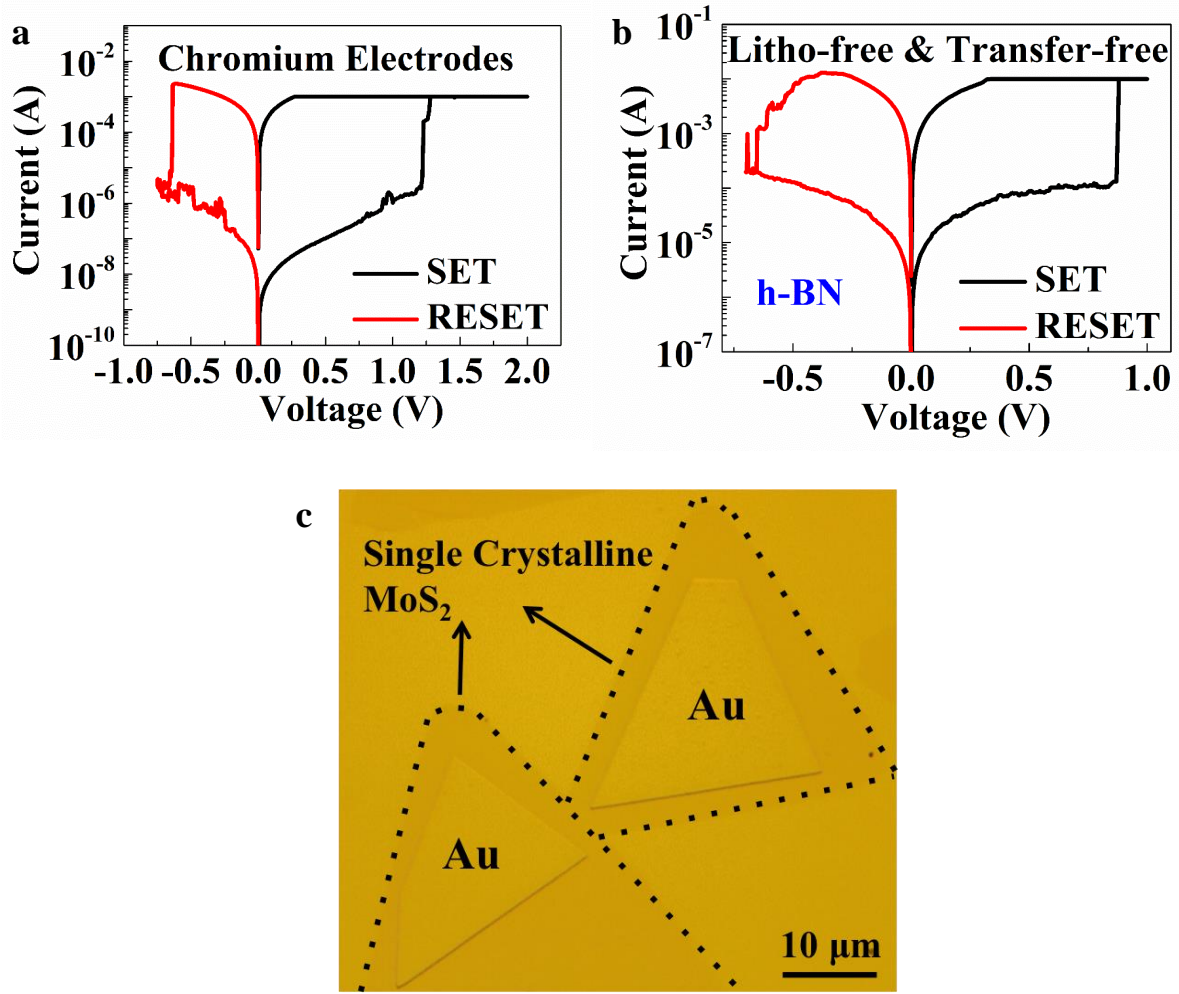

**Supplementary Figure 5.** Representative I-V curves and optical image of different MIM device structures. (a) Representative I-V curve of litho-free 4-layer MoS$_2$ memory devices using chromium as both top and bottom electrodes. The area of this Cr/MoS$_2$/Cr litho-free device is 15 × 15 μm$^2$ and HRS tunnel resistance is ~2.4 GΩ-μm$^2$. (b) Representative I-V curve of litho-free and transfer-free h-BN MIM devices using nickel foil as the bottom electrode and gold as the top electrode. The h-BN layer is grown on nickel foil, thus transfer process is unnecessary. The area of this litho-free device is 10 × 10 μm$^2$ and HRS tunnel resistance is ~0.6 MΩ-μm$^2$. (c) Optical image of single crystalline monolayer MoS$_2$ MIM devices with Au top

and bottom electrodes.

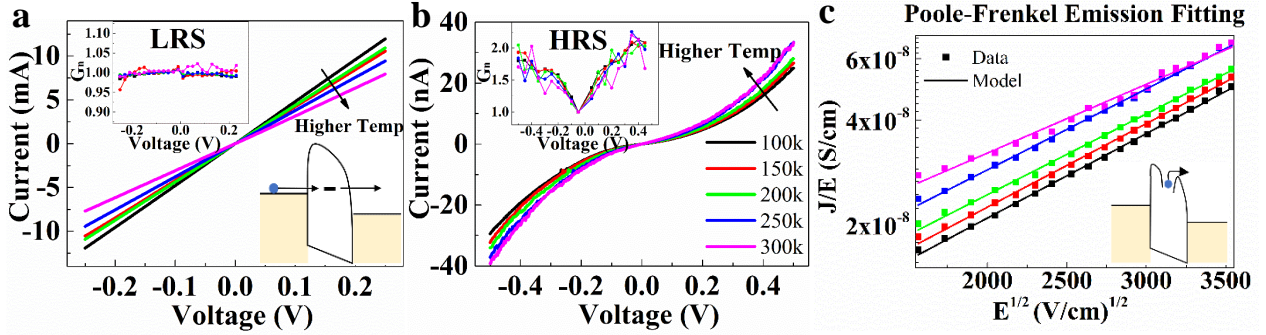

**Supplementary Figure 6. I-V characteristics of h-BN device at low temperatures.** (**a**) The temperature-dependent I-V characteristics at LRS based on h-BN crossbar devices indicating a metallic character since the current decreases with increasing temperature and conduction is linear. The inset shows the normalized conductance $G_n$ and the schematic of proposed mechanism based on Ohmic direct tunneling conduction via defect(s) at LRS. (**b**) The temperature-dependent I-V characteristics at HRS. Unlike the LRS, the current increases as the temperature increases in the HRS state. The inset shows the normalized conductance $G_n$, which is consistent with the Poole-Frenkel (P-F) emission model. (**c**) Fitted data using P-F emission model for HRS showing good agreement. The inset shows the schematic of the P-F model. The expression for Poole-Frenkel emission is $J \propto E \exp\left[\frac{-q(\phi_T - \sqrt{qE/\pi\varepsilon_r\varepsilon_0})}{kT}\right]$, where $J$ is the current density, T is the absolute temperature, q is the electronic charge, $q\phi_T$ is the trap energy level, $E$ is the electric field across the dielectric, k is Boltzmann's constant, $\varepsilon_0$ is the permittivity in vacuum, and $\varepsilon_r$ is the optical dielectric constant. In this case, the effective thickness is 0.4 nm and the device area is 2 x 2 μm$^2$. The temperature legend for all figures is shown in part b.

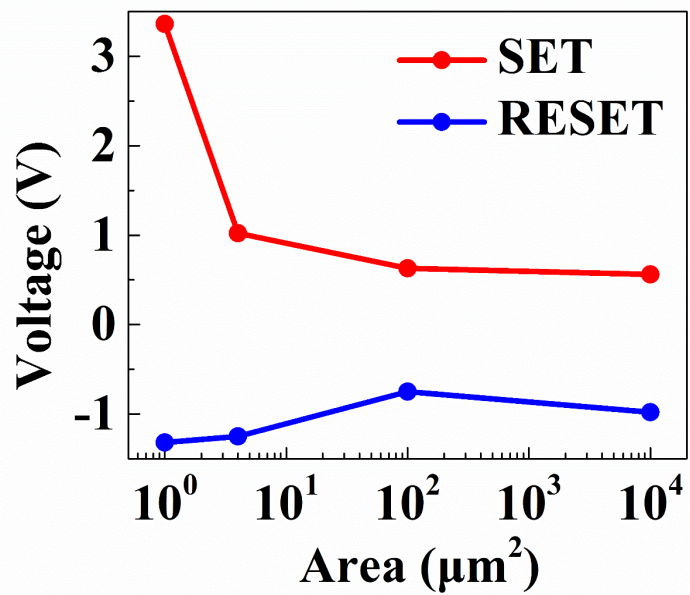

**Supplementary Figure 7.** Dependence of SET/RESET voltage on area. SET & RESET voltages vs. area for MoS$_2$ crossbar devices. As the device area increases, SET voltage decreases from 3 V to 0.5 V, while RESET voltage remains relatively flat around -1 V.

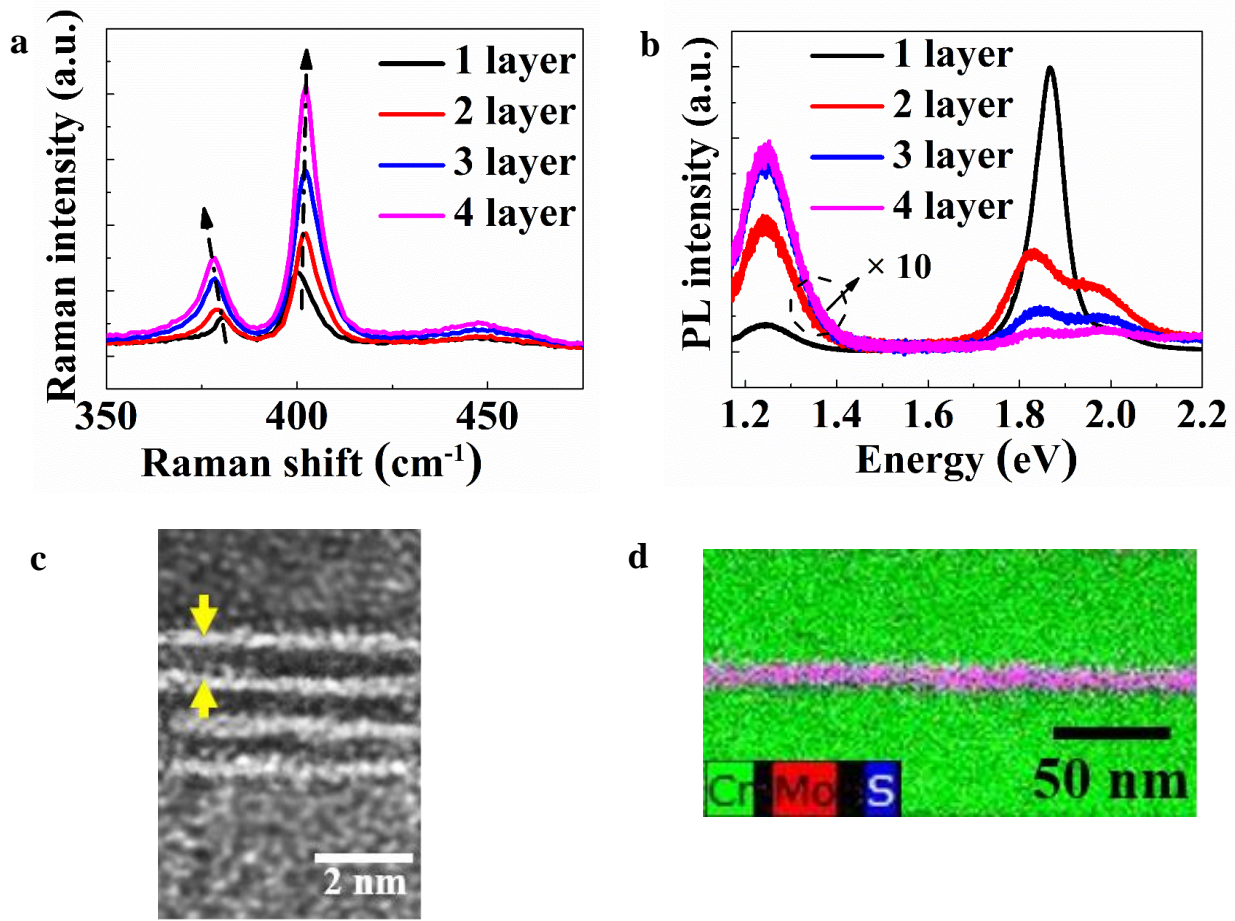

**Supplementary Figure 8. Characterizations of few-layer MoS$_2$.** (**a**) Raman spectra for different layers show the two most prominent peaks namely the E$^1_{2g}$ and A$_{1g}$ modes. The Raman shift of E$_{2g}$ mode decreases whereas that of A$_{1g}$ mode increases with increasing layer number. (**b**) Photoluminescence spectra showing decreasing intensity with increasing layer number. Intensity of 2-4 layer MoS$_2$ spectra is magnified by 10x for visibility. (**c**) TEM cross-section image of Cr/4 layer MoS$_2$/Cr litho-free device revealing the clear MoS$_2$ layer structure. The distance between the two arrows is 0.65 nm. (**d**) Elemental mapping of Cr/4-layer MoS$_2$/Cr litho-free device. The middle pink layer is the result of color combination of red Mo layer and blue S layer.

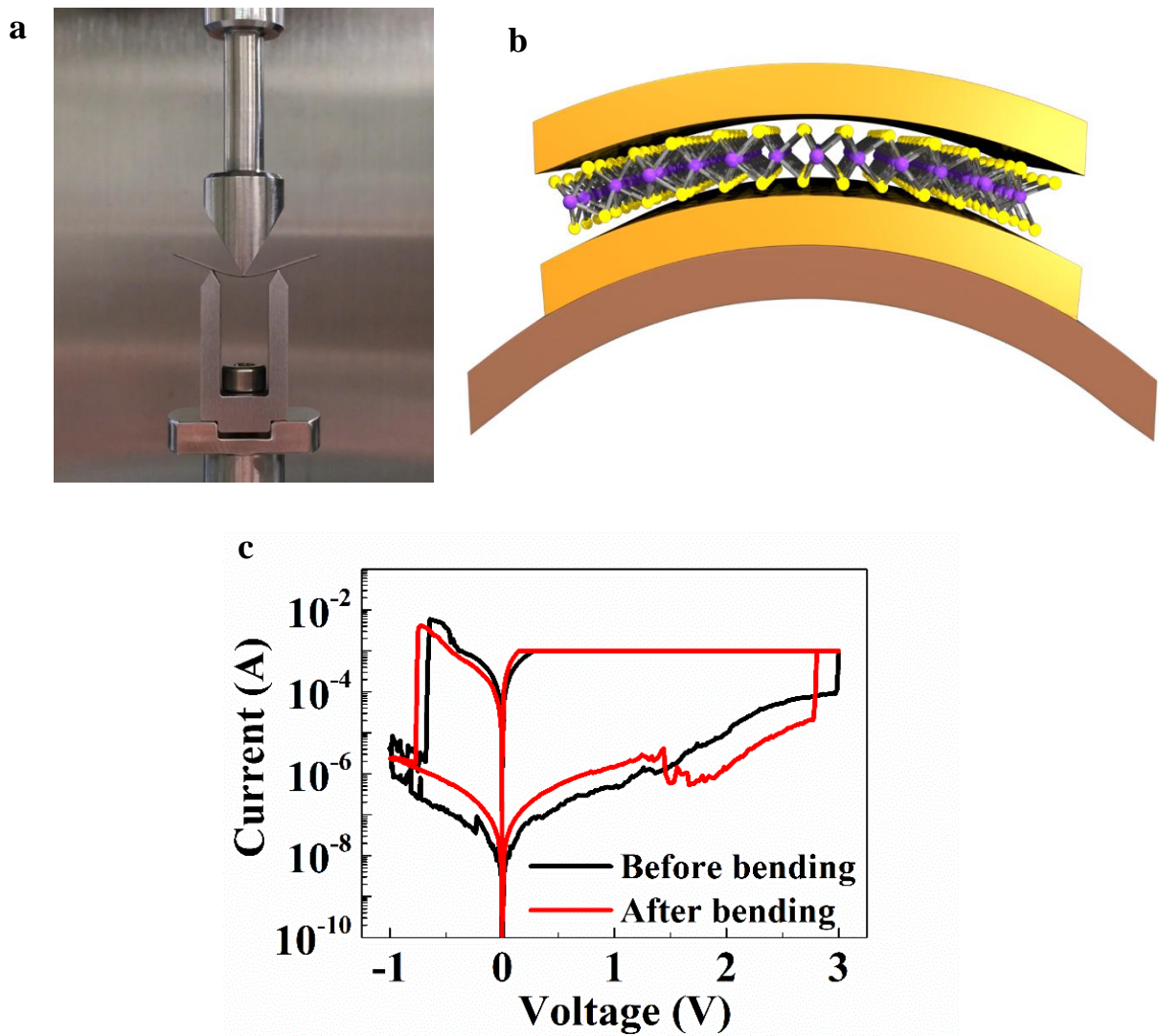

**Supplementary Figure 9.** DMA apparatus and switching I-V curves before and after bending. (a) Photography of the bent bilayer $MoS_2$ crossbar devices on PI substrate via DMA apparatus. (b) Schematic of the bent $MoS_2$ device sandwiched by gold electrodes on flexible substrate. (c) Typical switching I-V curves of bilayer $MoS_2$ crossbar devices before and after 1000 cycles at 1% strain. This device is the same one with the LRS device in Fig. 5e.

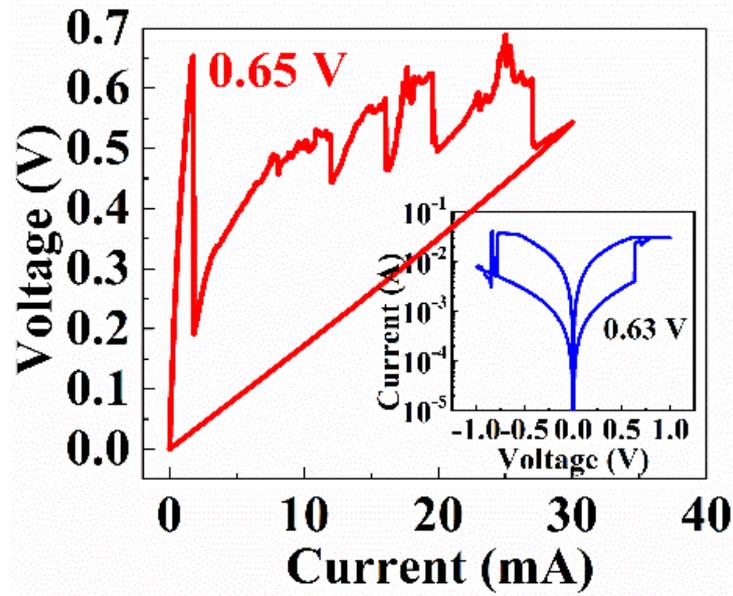

**Supplementary Figure 10.** Current sweep SET operation of monolayer $MoS_2$ crossbar device. The current increases from 0 to 30 mA and then decreases to 0. The voltage increases with current until the voltage reaches 0.65 V, and then suddenly drops to around 0.2 V, which indicates the first resistance decreasing process (SET) at 0.65 V. From 0 to 30 mA, it has five separate SET steps. As the current decreases from 30 to 0 mA, the voltage follows a linear decreasing step back to 0 V. The inset shows the voltage sweep in the same device, with the SET voltage to be 0.63 V, corresponding to the first transition in current sweep at 0.65 V. The area of this crossbar device is $10 \times 10$ μm$^2$.

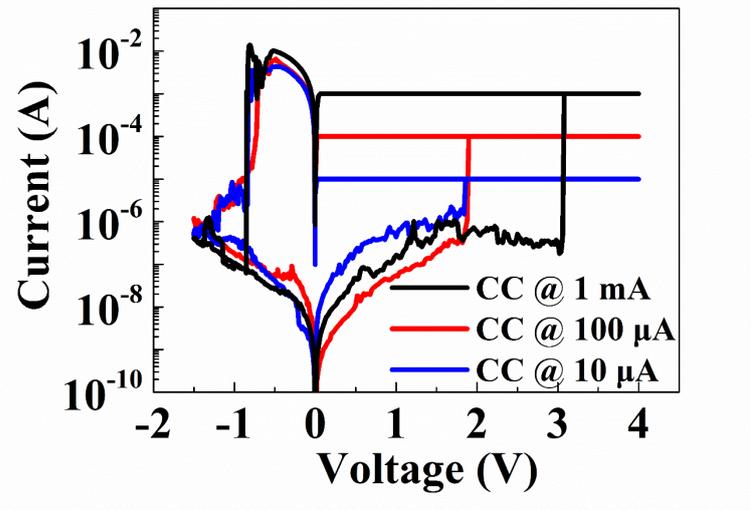

**Supplementary Figure 11.** Low compliance current operation of h-BN crossbar device. The compliance current during SET process is 1 mA, 100 μA and 10 μA respectively. The capability of operating at low compliance current shows the possibility of low-power non-volatile resistance switching. The area of the crossbar device is 2 × 2 μm$^2$.